\documentclass[aps,prb, twocolumn, reprint, floatfix,showpacs]{revtex4-1}

\usepackage{graphicx}
\usepackage{color}
\usepackage{amsmath}
\usepackage{amssymb}
\usepackage{amsthm}
\usepackage{hyperref}
\usepackage{soul}

\usepackage{graphicx}

\usepackage[usenames,dvipsnames]{xcolor}
\definecolor{tealgreen}{rgb}{0.0, 0.51, 0.5}
\definecolor{terracotta}{rgb}{0.89, 0.45, 0.36}
\definecolor{thulianpink}{rgb}{0.87, 0.44, 0.63}
\definecolor{trueblue}{rgb}{0.0, 0.45, 0.81}

\newcommand{\mathsym}[1]{{}}

\def  \bsig    {\mbox{\boldmath$\sigma$}}

\def  \bB      {\mbox{\boldmath$\mathcal{B}$}}
\def  \bA      {\mbox{\boldmath$\mathcal{A}$}}

\begin{document}

\title{Current-induced spin polarization in isotropic $k$-cubed Rashba model:
	 Theoretical study for p-doped semiconductor heterostructures and  perovskite oxides interfaces}
\author{{\L}. Karwacki$^{1}$, A. Dyrda\l$^{2,3}$, J. Berakdar$^{2}$, J. Barna{\'s}$^{1,3}$}
\affiliation{
$^{1}$Institute of Molecular Physics, Polish Academy of Sciences, ul. M. Smoluchowskiego 17, 60-179 Pozna\'n, Poland\\
$^{2}$ Institut f{\"u}r Physik, Martin-Luther Universit{\"a}t Halle-Wittenberg, D-06099 Halle, Germany\\
$^{3}$Faculty of Physics, Adam Mickiewicz University, ul. Umultowska 85, 61-614 Pozna\'n, Poland
}

%LaAlO$_{3}$/SrTiO$_{3}$

\date{\today }

\begin{abstract}
Using the Matsubara Green's function formalism we calculate the temperature dependence of the nonequilibrium spin polarization induced by an external electric field in the presence of spin-orbit coupling. The model Hamiltonian includes an isotropic $k$-cubed form of the Rashba spin-orbit interaction. Such a Hamiltonian captures the electronic and spin properties of two-dimensional  electron (hole) gas  at the surfaces or interfaces of transition metal oxides or in p-doped semiconductor heterostructures. The induced spin polarization is calculated for the nonmagnetic as well as magnetic electron/hole gas. Relation of the spin polarization to  the Berry curvature is also discussed.
\end{abstract}
\pacs{71.70.Ej, 85.75.-d, 72.25.Mk}

\maketitle

\section{Introduction}
The efficient control of electron spin is currently one of the key issues in spintronics. It is known that spin-orbit interaction in low dimensional systems is usually strongly enhanced and leads to new phases of matter that emerge at the interface~\cite{Fert2016}, like for instance chiral spin order, and to spin polarization. Thus, pure electrical control of the spin degree of freedom seems to be a promising concept for future applications in electronics. Moreover, such a control  is also intriguing from the fundamental physics point of view~\cite{Fert2016}. Indeed, the physics of low-dimensional heterostructures based on semiconductors, graphene-like materials, and oxides reveals a diversity of novel phenomena dictating  electronic and spin transport.

One of the most prominent phenomena induced by spin-orbit coupling is the spin Hall effect~\cite{hirsch,Dyakonov1971,Dyakonov1971_2}. This effect  has already become the standard tool for the generation and detection of spin currents~\cite{Ozyilmaz2004,Kiselev2003,Olejnik_2012_rev,Sinova2015_rev}. Spin Hall currents can furthermore generate spin-torque and inducing spin dynamics~\cite{Liu2011,MironGarello,LiuPai,CecotKarwacki}. In this scenario  one needs no magnetic polarizer as required for spin-valve devices.
The spin-orbit interaction may also  lead to spin polarization (a phenomenon  known as the Edelstein effect or the inverse spin galvanic effect) when an external electric field is applied to the system~\cite{dyakonov71b,Ivchenko78,edelstein90,aronov89}. In magnetic systems this non-equilibrium  spin polarization may interact via the exchange coupling with the local magnetization giving rise to a spin torque.

In the context of the aforementioned issues,  heterostructures of transition metal oxides are attracting much attention recently. Current experimental techniques allow for epitaxial growth of different high quality perovskite oxides as to  achieve artificially tailored heterostructures with relatively sharp interfaces~\cite{Ohtomo2002,Ohtomo2004a,Brinkam,Ohtomo2004b}. The discovery of two-dimensional (2D) electron gas at the interface of such structures like LaAlO$_{3}$/SrTiO$_{3}$ (LAO/STO)~\cite{Ohtomo2004a,Thiel}, offers a new route for materials and spintronics research, revealing a variety of  interesting phenomena in these hetrostructures -- starting from the colossal magnetoresistance, ferroelectricity, ferro- and antiferromagnetism, through metal-insulator transition and high temperature  superconductivity, and ending at large spin-orbit coupling~\cite   {CavigliaTriscone2008,CavigliaTriscone2010,Jackson,Kimura2003,Tokura,Glavic,Ruhman,Stornaiuolo,Biscaras,Bucheli,LesneFertVila,Schmidt2016,Raimondi2017,Cen,Richter,Bert}.

Despite the fact that cubic perovskites, such as STO and STO-based structures, have been studied intensively, there are only a few experimental results that reveal properties of their conduction bands. Moreover, the physical picture of the spin-orbit interaction in 2D electron gas at the interfaces of perovskite oxides, being under intensive discussion in recent years, is still elusive.

First of all, the STO-based heterostructures reveal  $d$-electron spin-orbit coupling~\cite{MacDonald2011,Khalsa2012,Khalsa2013,DiXiao2015}. The strongly anisotropic $d$-orbitals together with quantum confinement result in complicated spin-orbit texture that is much richer  than the case of  $sp$-electron gas in the conventional n-doped semiconductor heterostructures~\cite{winkler}. In case of cubic perovskites, such as STO, the crystal field due to the octahedral coordination with neighboring oxigen atoms splits the degenerate atomic d-levels (originating mainly from  Ti sites) into the threefold degenerate t$_{2g}$ states and twofold degenerate e$_{g}$ states~\cite{Khalsa2013,Fasolino2013,Fasolino2017}. The  energy distance between these orbitals is about 2 eV.  Therefore, low-energy effective models  of electronic states in the vicinity of $\Gamma$ point of the Brillouin zone take into account only the t$_{2g}$ orbitals. The symmetry of the bottom of the conduction  t$_{2g}$ band  (at the $\Gamma$ point) is the same as the symmetry of the corresponding $p$-states in the p-doped semiconductor heterostructures based on zincblende III-V semiconductors~\cite{Fasolino2017,Nakamura2012}. Accordingly, the spin-orbit coupling lifts the degeneracy at the $\Gamma$ point further into a heavy and light electron bands (with the total angular momentum $J = 3/2$) and split off band (with $J = 1/2$), similar to the heavy, light and split off hole bands in III-V semiconductors.

Recent experimental results based on weak localization and antilocalization effect in magnetoresistance  indicate unambiguously the k-cubed character of Rashba spin-orbit interaction
in transition metal oxides~\cite{Nakamura2012,LiangZhang2015}. This is in agreement with recent theoretical studies based on DFT simulations and tight-binding modelling~\cite{Fasolino2017,Zhong2013}.

In this paper, we study in details the current-induced spin polarization for the effective Hamiltonian describing 2D electron/hole gas with isotropic k-cubed Rashba spin-orbit interaction. The model Hamiltonian  under consideration has the form of a $2\times 2$ matrix, and has been derived by the perturbation procedure from the $8\times 8$ Luttinger Hamiltonian for p-doped semiconductor quantum wells with structural inversion asymmetry~\cite{LiuShenZhu2008}. Such a model was used to describe  the experimentally observed 2D electron gas at the oxides interfaces~\cite{Nakamura2012}. Therefore in this paper we focus on spin and electronic transport properties of electron and  hole gas, that in the first approximation can be described by the effective Hamiltonian with the isotropic k-cubed form of Rashba spin-orbit interaction. To describe the current-induced spin polarization we use  the Matsubara Green's function formalism in the linear response regime. This allows us to analyse the non-equilibrium spin polarization also beyond the zero-temperature limit.

The paper is organized as follows. In Sec. II. we introduce the effective Hamiltonian and  the necessary  concepts  for the analytical and numerical calculations.  In Sec. III. we discuss the current-induced spin polarization and its temperature dependence in a non-magnetic k-cubed Rashba gas. In Sec. IV. we present our results for the system in the presence of the exchange field. At first, we discuss some special cases,  where the magnetization is oriented  perpendicularly to the 2D gas plane and when it is oriented in plane of 2D gas. At the end of this section we also discuss the case of arbitrary oriented exchange field. The final conclusions and outlook for future research are presented in Sec. V.

\section{Theoretical outline}
\subsection{Model}
We consider  the effective Hamiltonian describing 2D electron (hole) gas with isotropic k-cubed Rashba spin-orbit interaction and subject to an exchange field. With  some assumptions such an effective Hamiltonian may be appropriate for description of two-dimensional electron gas (2DEG) at the interface between two oxide perovskites, like for instance LAO/STO~\cite{Nakamura2012}, or for heavy-hole gas that appears in  semiconductor heterostructures~\cite{winkler}. This
hamiltonian takes the following matrix form:
\begin{equation}\label{RisoM}
\hat{H} = \frac{\hbar^{2} k^{2}}{2 m} \sigma_{0} + \mathrm{i} \lambda \left( k_{-}^{3} \sigma_{+} - k_{+}^{3} \sigma_{-}\right) - \frac{1}{\hbar}
\mathbf{H} \cdot \hat{\mathbf{S}},
\end{equation}
where the first term describes the kinetic energy with the effective mass defined  by electron rest mass $m_{0}$ and Luttinger parameters $\gamma_{1,2}$~\cite{LiuShenZhu2008}:
\begin{equation}
m = m_{0} \left(\gamma_{1} + \gamma_{2} - \frac{256 \gamma_{2}^{2}}{3 \pi^{2} (3 \gamma_{1} + 10 \gamma_{2})}\right)^{-1} .
\end{equation}
The second term describes the isotropic k-cubed Rashba spin-orbit interaction with $k_{\pm} = k_{x} \pm \mathrm{i} k_{y}$, $\sigma_{\pm} = (\sigma_{x} \pm \mathrm{i} \sigma_{y})/2$ (here, $\sigma_{\alpha}$ with $\alpha = 0, x, y, z$ are the unit and Pauli matrices, respectively), and the Rashba coupling parameter defined as~\cite{LiuShenZhu2008}:
\begin{equation}
\label{lambda0}
\lambda = \frac{512 e F L_{z}^{4} \gamma_{2}^{2}}{9 \pi^{2} (3 \gamma_{1} + 10 \gamma_{2})(\gamma_{1} - 2 \gamma_{2})},
\end{equation}
\begin{figure}[t]
	\includegraphics[width=1.0\columnwidth]{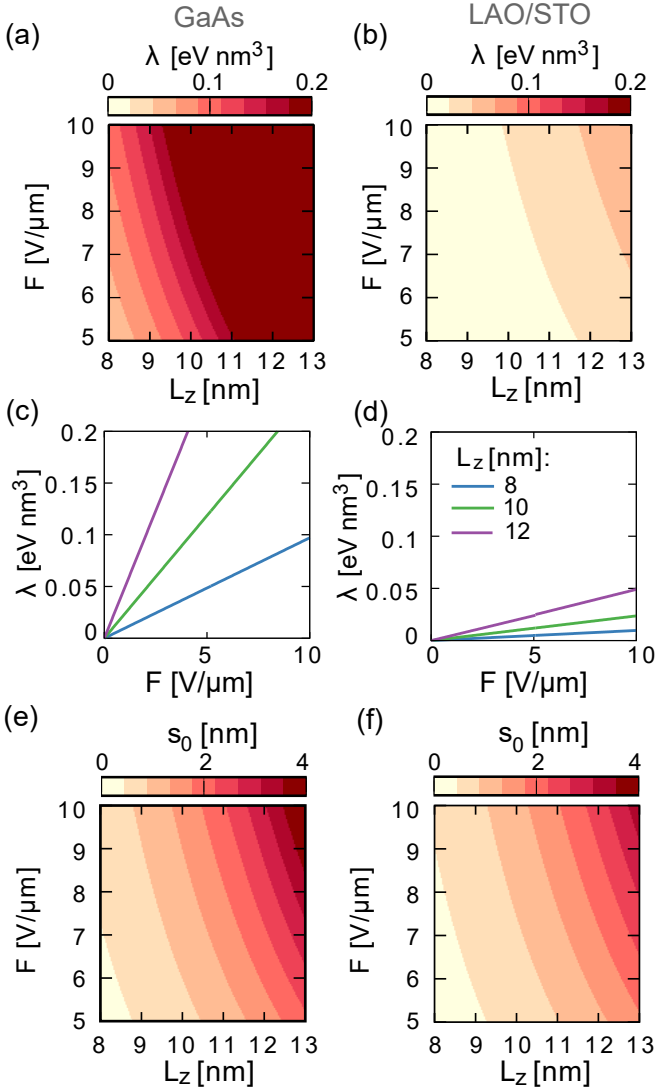}
	\caption{Spin-orbit coupling strength $\lambda$ as a function of quantum well width $L_z$ and confining electric field $F$ for GaAs (a) and perovskite oxide (b); cross-sections of $\lambda$ as functions of $F$ for GaAs (c) and perovskite oxide (d); $s_0$ parameter as a function of quantum well width $L_z$ and confining electric field $F$ for GaAs (e) and perovskite oxide (f). Luttinger parameters: (GaAs) $\gamma_1=6.85$, $\gamma_2=0.21$, (LAO/STO) $\gamma_1=2.3$, $\gamma_2=0.26$. }
\label{fig:fig1}
\end{figure}
where $L_{z}$ and $eF$ denote the width and the potential of the quantum well, respectively. The last term in  Hamiltonian (1) describes the effect of exchange field due to the exchange interaction between electrons and local macroscopic magnetization. The exchange field $\mathbf{H}$ is oriented arbitrarily and its three components in the spherical coordinate system can be written as follows:
\begin{eqnarray}
H_{x} = H_{0} \sin\theta \cos\xi , \\
H_{y} = H_{0} \sin\theta \sin\xi , \\
H_{z} = H_{0} \cos\theta ,
\end{eqnarray}
%where $H_{0} = h_0 M_{s} \left[ 1 - (T/T_{c})^{3/2}\right]$
where $H_{0} = h_{0} \left[ 1 - (T/T_{c})^{3/2}\right]$ with $h_{0}$ given in energy units and proportional to the exchange parameter and the saturation magnetization at $T = 0$, and $T_{c}$ denoting the Curie temperature.

The Hamiltonian (\ref{RisoM}) has been obtained upon two canonical transformations, so the spin-operators, $\hat{S}_{\alpha}$, after the same  unitary transformations have the form:
\begin{eqnarray}
\hat{S}_{x} =  - \hbar s_{0} k_{y} \sigma_{0} + s_{1} \hbar  (k_{-}^{2} \sigma_{+} + k_{+}^{2} \sigma_{-}) \hspace{1.5cm}\nonumber\\= - \hbar s_{0} k_{y} \sigma_{0} + \hbar s_{1} (k_{x}^{2} - k_{y}^{2}) \sigma_{x} + 2 \hbar s_{1} k_{x} k_{y} \sigma_{y},
\end{eqnarray}

\begin{eqnarray}
\hat{S}_{y} = s_{0} \hbar k_{x} \sigma_{0} + \mathrm{i} s_{1} \hbar  (k_{-}^{2} \sigma_{+} - k_{+}^{2} \sigma_{-}) \hspace{1.5cm} \nonumber \\
= \hbar s_{0} k_{x} \sigma_{0} + \hbar s_{1} (k_{x}^{2} - k_{y}^{2}) \sigma_{y} - 2 \hbar s_{1} k_{x} k_{y} \sigma_{x},
\end{eqnarray}

\begin{eqnarray}
\hat{S}_{z} = \frac{3}{2} \hbar \sigma_{z} ,
\end{eqnarray}
where $s_0$ and $s_{1}$ are defined by the material parameters:
\begin{equation}
s_{0} = \frac{512 \gamma_{2} L_{z}^{4} e F m_{0}}{9 \pi^{6} (3 \gamma_{1} + 10 \gamma_{2})(\gamma_{1} - 2 \gamma_{2}) \hbar^{2}},
\end{equation}
\begin{equation}
s_{1} = L_{z}^{2} \left(\frac{3}{4 \pi^{2}} - \frac{256 \gamma_{2}^{2}}{3 \pi^{4} (3 \gamma_{1} + 10 \gamma_{2})} \right),
\end{equation}
(for details see [\onlinecite{LiuShenZhu2008}]). Thus, the effective mass as well as the Rashba  parameter depend strongly on the material parameters. Variation of the Rashba parameter and the parameter $s_{0}$ with the quantum well width  $L_{z}$ and the electric field $F$ (describing the asymmetric quantum well potential) are shown in Fig. 1.  We present this dependence for Luttinger parameters $\gamma_{1,2}$ adequate for GaAs 2DHG~\cite{}  (Figs.~\ref{fig:fig1}(a),(c),(e)) and for $\gamma_{1,2}$ obtained from fitting to experimental data and from DFT calculations for the LAO/STO interface~\cite{Fasolino2013,Nakamura2012,LiangZhang2015} (Figs.~\ref{fig:fig1}(b),(d),(f)). Thus, for fixed Luttinger parameters the Rashba spin-orbit coupling increases with increasing electric field and width of the quantum well. Similar conclusion  follows from the behavior of the $s_0$ parameter, which is shown in Fig.~\ref{fig:fig1}(e) and (f). Moreover, for chosen values of Luttinger parameters, the spin-orbit interaction is stronger in  GaAs two-dimensional hole gas (2DHG) than in LAO/STO 2DEG. We should also note that the transformations leading to expression~(\ref{lambda0}) take into account only $\gamma_1$ and $\gamma_2$ parameters of the original Luttinger Hamiltonian. In the general case, however, an additional $\gamma_3$ parameter is present and may play a role when estimating the proper parameters of the system~\cite{winkler}.

\subsection{Method and general solution}
The nonequilibrium spin polarization created by the external electric field can be  calculated in the Matsubara-Green functions formalism from the formula~\cite{mahan}:
{\small{\begin{eqnarray}
		\label{S_alpha_Matsubara}
		S_{i} (\mathrm{i} \omega_{m})= \frac{1}{\beta}
		\sum_{\mathbf{k}, n} \mathrm{Tr}\left\{\hat{S}_{i} G_{\mathbf{k}}(\mathrm{i} \varepsilon_{n} + \mathrm{i} \omega_{m}) \hat{H}_{\mathbf{A}} (\mathrm{i} \omega_{m})G_{\mathbf{k}}(\mathrm{i} \varepsilon_{n}) \right\}, \hspace{0.5cm}
		\end{eqnarray}}}
where $\beta =1/k_BT$ (with $T$ and $k_B$ denoting the temperature and Boltzmann constant, respectively), $\varepsilon_{n}=(2n+1)\pi k_BT$ and $\omega_{m}=2m\pi k_BT$ are the Matsubara energies, while  $G_{\mathbf{k}}(\mathrm{i} \varepsilon_{n})$ are the Matsubara Green functions (in the $2\times 2$ matrix form).
The  perturbation term describing interaction of the system with an external electric field takes the form $\hat{H}_{\mathbf{A}}^{\scriptstyle{E}}(\mathrm{i} \omega_{m})=-e\hat{v}_jA_j(\mathrm{i} \omega_{m})$, with the amplitude of the vector potential $A_{j}(\mathrm{i} \omega_{m})$ determined by the amplitude $E_j(\mathrm{i} \omega_{m})$ of electric field through the following relation $A_{j}(\mathrm{i} \omega_{m}) = \frac{E_{j}(\mathrm{i} \omega_{m}) \hbar}{\mathrm{i} (\mathrm{i} \omega_{m})}$.

 The summation over Matsubara frequencies in Eq.(\ref{S_alpha_Matsubara}) can be performed using  contour integration ~\cite{mahan}:
\begin{eqnarray}
\frac{1}{\beta} \sum_{n} \hat{S}_{i} G_{\mathbf{k}}(\mathrm{i} \varepsilon_{n} + \mathrm{i} \omega_{m}) \hat{v}_{j} G_{\mathbf{k}}(\mathrm{i} \varepsilon_{n})\hspace{2cm}\nonumber\\ = - \int_{\mathcal{C}} \frac{dz}{2\pi \mathrm{i}} f(z) \hat{S}_{i} G_{\mathbf{k}}(z + \mathrm{i} \omega_{m}) \hat{v}_{j} G_{\mathbf{k}}(z),
\end{eqnarray}
where $f(z)$ is the meromorphic function of the form $[\exp(\beta z) + 1]^{-1}$, which
has simple poles at the odd integers $n$ ($z = \mathrm{i}\varepsilon_n$) and $\mathcal{C}$ is an appropriate integration contour. After an  analytical continuation~\cite{mahan} we find the general expression describing nonequilibrium spin polarization in the following form~\cite{Dyrdal2017}:
\begin{eqnarray}
\label{SalphaE_final}
S_{i}(\omega) = - \frac{e \hbar}{\omega} E_{j}  \hspace{6.7cm}\nonumber \\
\times {\mathrm{Tr}} \int \frac{d^{2}\mathbf{k}}{(2\pi)^{2}} \int \frac{d \varepsilon}{2 \pi} f(\varepsilon) \hat{S}_{i} \Bigl( G_{\mathbf{k}}^{R}(\varepsilon + \omega) \hat{v}_{j} [G_{\mathbf{k}}^{R}(\varepsilon) - G_{\mathbf{k}}^{A}(\varepsilon)]\Bigr. \hspace{0.7cm}\nonumber\\
+ \Bigl. [G_{\mathbf{k}}^{R}(\varepsilon) - G_{\mathbf{k}}^{A}(\varepsilon)] \hat{v}_{j} G_{\mathbf{k}}^{A}(\varepsilon - \omega)\Bigr), \hspace{0.7cm}
\end{eqnarray}
where $f(\varepsilon)$ is the Fermi-Dirac distribution function and $\hat{v}_{j}$ is the $j$-th component of the velocity operator,  $\hat{v}_{j} = (1/\hbar )\partial \hat{H}/\partial k_{j} $.

To make our further expressions more clear let us rewrite the Hamiltonian (\ref{RisoM}) in the general form
 \begin{equation}
 H = n_{0} \sigma_{0} + \mathbf{n} \cdot \bsig ,
 \end{equation}
 where  $\mathbf{n} = (n_{x}, n_{y}, n_{z})$ and $\bsig = (\sigma_{x}, \sigma_{y}, \sigma_{z})$. The coefficients $n_{i}$ ($i = 0, x, y, z$) take then the following forms:

 \begin{eqnarray}
 n_{0} = \varepsilon_k - s_{0} (k_{y} H_{x} - k_{x} H_{y}),\hspace{4.45cm}\\
 n_{x} = - \lambda (k_{y}^{3} - 3 k_{x}^{2} k_{y}) - s_{1} H_{x} (k_{x}^{2} - k_{y}^{2}) - 2 H_{y} s_{1} k_{x} k_{y},\hspace{0.7cm}\\
 n_{y} = - \lambda (k_{x}^{3} - 3 k_{x} k_{y}^{2}) - s_{1} H_{y} (k_{x}^{2} - k_{y}^{2}) + 2 H_{x} s_{1} k_{x} k_{y},\hspace{0.7cm}\\
 n_{z} = - \frac{3}{2} H_{z}. \hspace{6.85cm}
 \end{eqnarray}

 The retarded Green function may then be written as:
 \begin{equation}
 \label{GF}
 G_{\mathbf{k}}^{R}(\varepsilon) =  G_{\mathbf{k} 0}^{R} \sigma_{0} + G_{\mathbf{k} x}^{R} \sigma_{x} + G_{\mathbf{k} y}^{R} \sigma_{y} + G_{\mathbf{k} z}^{R} \sigma_{z},
 \end{equation}
 with the coefficients:
 \begin{subequations}
 	\begin{align}
 G_{\mathbf{k} 0}^{R} &= \frac{1}{2} (G_{k +}^{R} + G_{k -}^{R}),\hspace{0.2cm}\\
 G_{\mathbf{k} x}^{R} &= \frac{n_{x}}{2 n} (G_{k +}^{R} - G_{k -}^{R}),\\
 G_{\mathbf{k} y}^{R} &= \frac{n_{y}}{2 n}  (G_{k +}^{R} - G_{k -}^{R}),\\
 G_{\mathbf{k} z}^{R} &= \frac{n_{z}}{2 n}  (G_{k +}^{R} - G_{k -}^{R}).
 \end{align}
 \end{subequations}
 Now $G_{k \pm}^{R} = [\varepsilon + \mu - E_{\pm} + \mathrm{i} \Gamma {\mathrm{sgn}}(\varepsilon)]^{-1}$ are defined by the eigenvalues $E_{\pm} = n_{0} \pm n$, and $n =\sqrt{n_{x}^{2} + n_{y}^{2} + n_{z}^{2}}$. The $i$-th component of the velocity operator is now given by the expression:
 \begin{equation}
 \label{vi}
 \hat{v}_{i} = \sum_{j = 0, x,y,z} \frac{1}{\hbar} \frac{\partial n_{j}}{\partial k_{i}} \sigma_{j}  \equiv \sum_{j} v_{ij} \sigma_{j} .
 %= v_{i0} \sigma_{0} + \mathbf{v}_{i} \cdot \bsig
 \end{equation}
 We also introduce the general form for the spin operator components:
 \begin{equation}
 \label{si}
 \hat{S}_{i} = \sum_{j = 0, x, y, z} s_{i j} \sigma_{j}.
 \end{equation}
Combining expressions (\ref{GF}) - (\ref{si}) with Eq.(\ref{SalphaE_final}) and performing the trace we obtain the following expressions for the components of current-induced spin polarization:
{\small{
\begin{eqnarray}
\label{sx}
S_{x}(\omega) = \frac{e \hbar}{\omega} E_{y} \int \frac{d^{2} \mathbf{k}}{(2\pi)^{2}} \Bigl\{\left[s_{x0} v_{y0} + s_{xx} v_{yx} \right] \mathcal{S}_{A}(\omega)\Bigr.
-  s_{xy} v_{yy} \mathcal{S}_{B}(\omega) \nonumber\\
-   \left[ \frac{n_{x}}{n} (s_{xx} v_{y0} + s_{x0} v_{yx})
+ \frac{n_{y}}{n} (s_{xy} v_{y0} + s_{x0} v_{yy})\right] \mathcal{S}_{C}(\omega)\nonumber\\
- \mathrm{i}  \frac{n_{z}}{n} \left[ s_{xy} v_{yx} - s_{xx} v_{yy}\right] \mathcal{S}_{D}(\omega) \nonumber\\
-  \frac{1}{n^{2}} \left[ (n_{y}^{2} + n_{z}^{2}) s_{xx} v_{yx} - n_{y}^{2} s_{xy} v_{yy}
\right. \nonumber\\\left.
 - \Bigl. n_{x} n_{y} (s_{xy} v_{yx} + s_{xx} v_{yy} )  \right] \mathcal{S}_{E}(\omega)\Bigr\},\nonumber\\
\end{eqnarray}
}}
{\small{
\begin{eqnarray}
\label{sy}
S_{y}(\omega) = \frac{e \hbar}{\omega} E_{y} \int \frac{d^{2} \mathbf{k}}{(2\pi)^{2}} \Bigl\{ \left[s_{y0} v_{y0} + s_{yx} v_{yx} \right] \mathcal{S}_{A}(\omega) \Bigr.
%\nonumber\\
-  s_{yy} v_{yy} \mathcal{S}_{B}(\omega) \nonumber\\
- \left[ \frac{n_{x}}{n} (s_{yx} v_{y0} + s_{y0} v_{yx})
+\frac{n_{y}}{n} (s_{yy} v_{y0} + s_{y0} v_{yy})\right] \mathcal{S}_{C}(\omega)\nonumber\\
- \mathrm{i}  \frac{n_{z}}{n} \left[ s_{yy} v_{yx} - s_{yx} v_{yy}\right] \mathcal{S}_{D}(\omega) \nonumber\\
-  \frac{1}{n^{2}} \left[ (n_{y}^{2} + n_{z}^{2}) s_{yx} v_{yx} - n_{y}^{2} s_{yy} v_{yy} \right. \nonumber\\
\left. - \Bigl.n_{x} n_{y} (s_{yy} v_{yx} + s_{yx} v_{yy} )  \right] \mathcal{S}_{E}(\omega)\Bigr\}, \nonumber\\
\end{eqnarray}
}}
{\small{
\begin{eqnarray}
\label{sz}
S_{z}(\omega) =
- \frac{e \hbar}{\omega} E_{y} \int \frac{d^{2} \mathbf{k}}{(2\pi)^{2}} \Bigl\{
\frac{n_{z}}{n} s_{zz} v_{y0} \mathcal{S}_{C}(\omega) \Bigr.\nonumber\\
- \mathrm{i}  \frac{s_{zz}}{n} \left[ n_{x} v_{yy} - n_{y} v_{yx}\right] \mathcal{S}_{D}(\omega)
\nonumber\\
-  \Bigl.\frac{n_{z}}{n^{2}} s_{zz} \left[ n_{x} v_{yx} + n_{y} v_{yy}\right] \mathcal{S}_{E}(\omega)\Bigr\},
\end{eqnarray}
}}
where the functions $\mathcal{S}_{A}$ to $\mathcal{S}_{E}$ have the  form:
\begin{eqnarray}
\mathcal{S}_{A} = I_{--}^{RA}(\omega) - I_{--}^{RR}(\omega) + I_{++}^{RA}(\omega) - I_{++}^{RR}(\omega)\hspace{0.8cm}\nonumber\\
+ I_{++}^{AA}(-\omega)- I_{--}^{RA}(-\omega) - I_{++}^{RA}(-\omega) + I_{--}^{AA}(-\omega),\hspace{0.13cm} \\
\mathcal{S}_{B} = I_{-+}^{RR}(\omega) - I_{-+}^{RA}(\omega) - I_{+-}^{RA}(\omega) + I_{+-}^{RR}(\omega)\hspace{0.8cm}\nonumber\\
+ I_{-+}^{RA}(-\omega) - I_{+-}^{AA}(-\omega) + I_{+-}^{RA}(-\omega) - I_{-+}^{AA}(-\omega), \hspace{0.1cm}\\
 \mathcal{S}_{C} = I_{--}^{RA}(\omega) - I_{--}^{RR}(\omega) - I_{++}^{RA}(\omega) + I_{++}^{RR}(\omega)\hspace{0.8cm}\nonumber\\
+ I_{--}^{AA}(-\omega)- I_{--}^{RA}(-\omega) + I_{++}^{RA}(-\omega) - I_{++}^{AA}(-\omega), \hspace{0.1cm}\\
\mathcal{S}_{D} = I_{-+}^{RA}(\omega) - I_{-+}^{RR}(\omega) - I_{+-}^{RA}(\omega) + I_{+-}^{RR}(\omega)\hspace{0.8cm}\nonumber\\
- I_{+-}^{AA}(-\omega) + I_{-+}^{AA}(-\omega) - I_{-+}^{RA}(-\omega) + I_{+-}^{RA}(-\omega),\hspace{0.1cm}\\
\mathcal{S}_{E} = I_{--}^{RA}(\omega) - I_{-+}^{RA}(\omega) - I_{--}^{RR}(\omega) + I_{-+}^{RR}(\omega) \hspace{0.8cm}\nonumber\\
- I_{+-}^{RA}(\omega) + I_{++}^{RA}(\omega) + I_{+-}^{RR}(\omega) - I_{++}^{RR}(-\omega)\nonumber\\
-I_{-+}^{AA}(-\omega) + I_{--}^{AA}(-\omega) - I_{+-}^{AA}(-\omega) + I_{++}^{AA}(-\omega)\nonumber\\
- I_{--}^{RA}(-\omega) + I_{-+}^{RA}(-\omega) + I_{+-}^{RA}(-\omega) - I_{++}^{RA}(-\omega),\hspace{0.1cm}
\end{eqnarray}
and each  $I_{\alpha \beta}^{XY}$ denotes integral over $\varepsilon$ defined as:  $I_{\alpha \beta}^{XY}(\omega) = \int \frac{d\varepsilon}{2\pi} f(\varepsilon) G_{\alpha}^{X}(\varepsilon+\omega) G_{\beta}^{Y}(\varepsilon)$ and $I_{\alpha \beta}^{XY}(-\omega) = \int \frac{d\varepsilon}{2\pi} f(\varepsilon) G_{\alpha}^{X}(\varepsilon) G_{\beta}^{Y}(\varepsilon - \omega)$. These integrals in their general forms have been derived in Refs.~[\onlinecite{Dyrdal2017}] and ~[\onlinecite{Dyrdal2016}].

\section{Nonmagnetic case}

In this section we revisit the model of nonmagnetic gas with isotropic k-cubed form of Rashba spin-orbit interaction. In such a case the Hamiltonian (\ref{RisoM}) reduces to the following form:
\begin{equation}\label{Riso}
H = \frac{\hbar^{2} k^{2}}{2 m} \sigma_{0} + \mathrm{i} \lambda \left( k_{-}^{3} \sigma_{+} - k_{+}^{3} \sigma_{-}\right),
\end{equation}
and the retarded Green's function corresponding to the Hamiltonian (\ref{Riso}) can be presented as
\begin{equation}
\label{G}
G_{\mathbf{k}}^{R}(\varepsilon)  = G_{\mathbf{k} 0}^{R} \sigma_{0} + G_{\mathbf{k} x}^{R} \sigma_{x} + G_{\mathbf{k} y}^{R} \sigma_{y}
\end{equation}
with the coefficients:
\begin{eqnarray}
G_{\mathbf{k} 0}^{R} = \frac{1}{2} (G_{k +}^{R} + G_{k -}^{R}),\hspace{1.2cm}\\
G_{\mathbf{k} x}^{R} = \sin(3\phi) (G_{k +}^{R} - G_{k -}^{R}),\hspace{0.4cm}\\
G_{\mathbf{k} y}^{R} = - \cos(3\phi) (G_{k +}^{R} - G_{k -}^{R}),
\end{eqnarray}
where  $E_{\pm} = \frac{\hbar^{2} k^{2}}{2m} \pm \lambda k^{3}$ denote the energy eigenvalues and $G_{k \pm}^{R} = [\varepsilon + \mu - E_{\pm} + \mathrm{i} \Gamma {\mathrm{sgn}}(\varepsilon)]^{-1}$. This model has been discussed in the literature in the context of spin Hall effect~\cite{SchliemannLoss} and current-induced spin polarization~\cite{LiuShenZhu2008} for 2D hole gas in the zero-temperature limit.

\begin{figure}[t]
	\includegraphics[width=1.0\columnwidth]{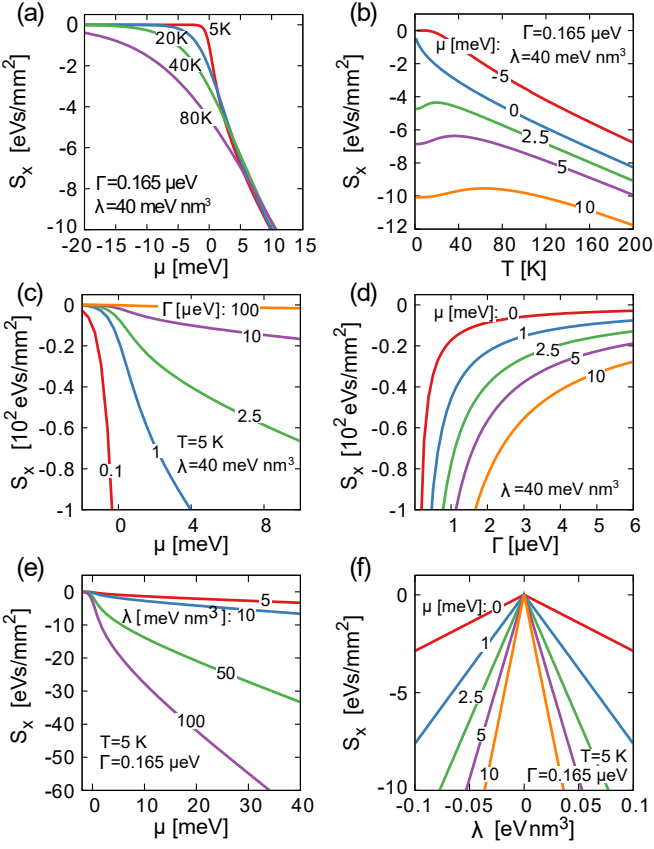}
	\caption{Current-induced spin polarization in the nonmagnetic case ($h_0=0$) as a function of chemical potential $\mu$ (a), (c), (e); temperature $T$ (b), relaxation rate $\Gamma$ (d) and spin-orbit coupling parameter $\lambda$ (f) for fixed parameters as indicated. The external electric field and the effective mass are choosen as $E_{y} = 1$eV/mm and $m = 0.12 m_{0}$, while the Luttinger parameters: $\gamma_{1} = 7$ and $\gamma_{2} = 0.27 \gamma_{1}$.  Other parameters (unless otherwise specified): $\lambda=0.04~\text{eV nm}^3$, $T=5~\text{K}$, and $\Gamma=1.65\cdot10^{-2}~\text{meV}$. Cut-off for integration over $\mathbf{k}$-vector has been assumed as $k_c=k_0/2$, where $k_0=\hbar^2/3m\lambda$~[\onlinecite{SchliemannLoss}].}
\label{fig:fig2}
\end{figure}

In the nonmagnetic case only the $x$-component of spin polarization is nonzero and Eq.(\ref{sx}) takes the following form in the dc-limit:
\begin{eqnarray}
\label{sxH0}
S_{x} =  e \hbar E_y \frac{s_{0}}{4 \pi} \left[3 \lambda \int \frac{dk k^{4}}{2\Gamma} [f'(E_{+}) - f'(E_{-})]\right.\nonumber\\
\left. + \frac{\hbar^{2}}{m} \int \frac{dk k^{3}}{2\Gamma} [f'(E_{+}) - f'(E_{-})] \right]\nonumber\\
-  e \hbar E_y \frac{s_{1}}{4\pi} \left[3 \lambda \int\frac{dk k^{5}}{2\Gamma}\bigg( \frac{f'(E_{+}) + f'(E_{-})}{1 + (\lambda k^{3}/\Gamma)^{2}}\right.\bigg.\nonumber\\
+ \bigg.f'(E_{+}) + f'(E_{-}) \bigg)\nonumber\\
\bigg. + \frac{\hbar^{2}}{m} \int \frac{dk k^{4}}{2 \Gamma} [f'(E_{+}) - f'(E_{-})]\bigg].
\end{eqnarray}

The integrals over $k$ in the expression above have analytical solutions in the low-temperature limit and lead to the following expression:
\begin{eqnarray}
S_{x} = - \frac{e E_y}{4 \Gamma} \hbar s_{0} \left[ 3 \lambda (k_{+}^{3} \nu_{+} - k_{-}^{3} \nu_{-}) + \frac{\hbar^{2}}{m} (k_{+}^{2} \nu_{+} + k_{-}^{2} \nu_{-})\right]\nonumber\\
+ \frac{e E_y}{4 \Gamma} \hbar s_{1} \left[ 3 \lambda  (k_{+}^{4} \nu_{+} + k_{-}^{4} \nu_{-}) + \frac{\hbar^{2}}{m}  (k_{+}^{3} \nu_{+} - k_{-}^{3} \nu_{-}) \right.\nonumber\\
+ \left. 3 \lambda \left( \frac{k_{+}^{4} \nu_{+}}{1 + (\lambda k_{+}^{3}/\Gamma)^{2}} + \frac{k_{-}^{4} \nu_{-}}{1 + (\lambda k_{-}^{3}/\Gamma)^{2}}\right)\right],\hspace{0.6cm}
\end{eqnarray}
where $k_{\pm}$ and $\nu_{\pm}$ are the Fermi wavevectors and densities of states corresponding to the $E_{\pm}$ energy subbands, respectively. The Fermi wavevectors are connected with the chemical potential $\mu$ and electron (hole) density $n$ by the following relations: $\mu = E_{+}(k_+)=E_{-}(k_-)$ and $n = (k_{+}^{2} + k_{-}^{2})/4\pi$. Thus, after some algebraic transformations one finds:
\begin{eqnarray}
\label{Sx_anal.}
S_{x} = - \frac{e E_y}{4 \Gamma} \hbar s_{0} \left[ 2 \mu (\nu_{+} + \nu_{-}) + \lambda(k_{+}^{3} \nu_{+} - k_{-}^{3} \nu_{-})\right]\nonumber\\
+\frac{e E_y}{4 \Gamma} \hbar s_{1} \left[  2 \mu  ( k_{+}\nu_{+} - k_{-} \nu_{-}) + \lambda(k_{+}^{4} \nu_{+} + k_{-}^{4} \nu_{-})\right.\nonumber\\
\left. + 3 \lambda \left( \frac{k_{+}^{4} \nu_{+}}{1 + (\lambda k_{+}^{3}/\Gamma)^{2}} + \frac{k_{-}^{4} \nu_{-}}{1 + (\lambda k_{-}^{3}/\Gamma)^{2}}\right)\right].
\end{eqnarray}

Equation (\ref{Sx_anal.}) may be treated as a counterpart of the Edelstein result for linear Rashba model~\cite{edelstein90}.  As the main contribution to the CISP is determined by the diagonal matrix elements of spin operators (proportional to $s_{0}$), the leading term in the equation above is the first one. Thus, similarly as in the case of $k$-linear Rashba coupling, the external electric field applied to the system induces non-equilibrium spin polarization which is aligned in plane of the two-dimensional gas and perpendicular to the external electric field. In both cases we also observe linear dependence on the relaxation time and on the Rashba coupling constant. The main difference between the k-linear and k-cubed Rashba models appears in their dependence  on the chemical potential. The Edelstein formula does not depend on the chemical potential, whereas the zero-temperature spin polarization for cubic Rashba model depends linearly on $\mu$.

Note that the above results have been obtained in the single loop approximation. However, it was reported that for randomly distributed point-like scatterers, the impurity vertex correction does not provide additional contribution to the transport properties -- in other words, the vertex correction to the velocity operator vanishes in this model~\cite{Murakami2004}. Furthermore, the relaxation rate $\Gamma$ (obtained as an imaginary part of the self-energy in Born approximation) is the same for both subbands.

Numerical results corresponding to Eq.(\ref{sxH0}) are presented in Fig.\ref{fig:fig2}.  Figures~\ref{fig:fig2}(a) and (b) show the temperature behavior of the spin polarization. Since the temperature leads to smearing of the carrier distribution in both subbands, one observes a nonzero spin polarization for negative chemical potentials. Note, that in our approach the chemical potential is fixed while the number of particle can vary. Moreover, spin polarization increases with increasing chemical potential. In a broad range of chemical potentials, this dependence is linear  with $\mu$ (see Fig.~\ref{fig:fig2}(a) and (c), (e)). Figures~\ref{fig:fig2} (c) and (d) highlight the dependence of spin polarization on the relaxation rate $\Gamma$. These plots clearly show a fast decrease of the spin polarization with increasing $\Gamma$. This decrease, however, depends strongly on the position of the Fermi level, which means that destructive effects associated with scattering on impurities may be slightly tuned by doping/gating of the system.
Finally, as the current-induced spin polarization considered here is driven by the spin-orbit interaction, it vanishes for $\lambda=0$, as shown in Figs.~\ref{fig:fig2}(e) and (f). The spin polarization also depends linearly on $\lambda$ as the difference between $E_-$ and $E_+$ bands changes linearly with $\lambda$ for a fixed Fermi level.

\begin{figure}[h]
	\includegraphics[width=1.0\columnwidth]{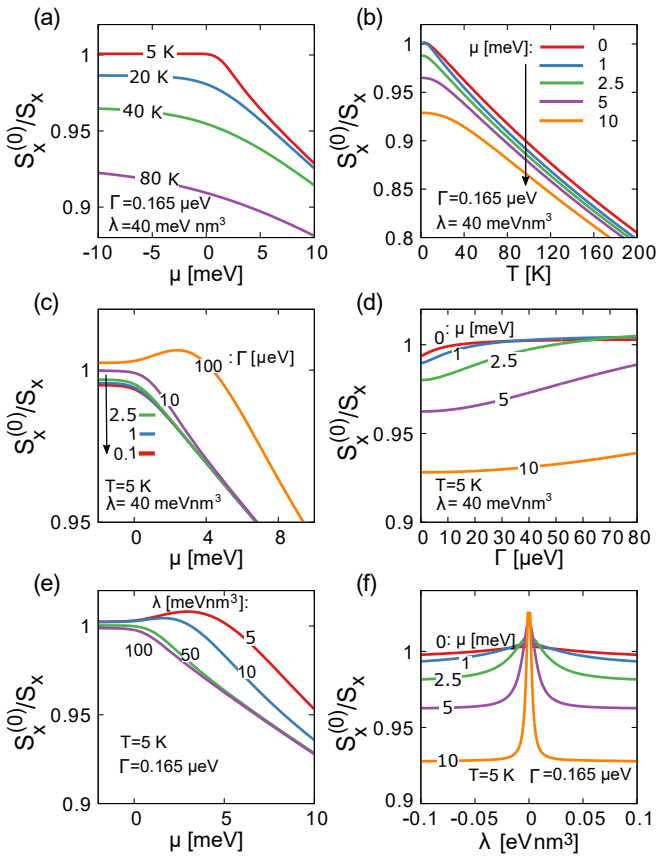}
	\caption{The ratio of $S_{x}(s_{1} = 0)/S_{x}(s_{1} \neq 0)=S_x^{(0)}/S_x$ in the nonmagnetic case. All parameters are taken as in Fig.2. }
	\label{fig:fig3}
\end{figure}
In our calculations we have included the terms proportional to $s_0$ and $s_1$. The latter  was neglected in previous studies~\cite{LiuShenZhu2008}. However, from our analysis follows that the term proportional to $s_1$ plays a remarkable role and should be included. In Fig.\ref{fig:fig3} we show the ratio of spin polarization calculated without (i.e. $S_{x}(s_1=0) = S_x^{(0)}$), and with the terms proportional to $s_1$ taken into account. Indeed, this figure shows that the term proportional to $s_1$ can lead to a correction of an order of up to 10\% or even larger at higher temperatures and larger Fermi levels, as shown in Figs.~\ref{fig:fig3}(a) and (b).
When the temperature increases,
  the correction to spin polarization due to the terms proportional to $s_1$ decreases with decreasing $T$ for small values of the chemical potential $\mu$, so the corresponding ratio $S_x^{(0)}/S_x$ becomes close to 1, see Fig.~3(b). In turn, the ratio $S_{x}^{(0)}/S_{x}$ only weakly depends on the impurity scattering rate $\Gamma$, as follows from see Fig.~3(d).
  When the Rashba spin-orbit coupling is weak, $S_x^{(0)}$ becomes larger than $S_x$, i.e. the contribution from the terms proportional to $s_1$  has opposite sign to that from terms proportional to  $s_0$ and the ratio $S_x^{(0)}/S_x$  exceeds 1, see
 Fig.~\ref{fig:fig3}(f). Similar situation also happens for large values of the chemical potentials, see  Fig.~\ref{fig:fig3}(e).
 When the Rashba parameter $\lambda$ increases, one observes a rapid decrease in  $S_{x}^{(0)}/S_{x}$ until this ratio  saturates at a certain level, see Fig.3(f).

\section{Magnetized 2D $k$-cubed Rashba gas}

In this section we consider the general case, when the time-reversal symmetry of the system is broken by the effective exchange field (see Hamiltonian 1).
Since the spin-orbit torques play an important role in various spintronics devices, we will analyze a general solution for an arbitrary oriented exchange field. Such a solution allows one to determine the spin-orbit torque induced by electric field in the system under consideration. Before this, however, we consider  two special cases -- when the exchange field is oriented perpendicularly to plane and when the exchange field is oriented in plane of 2D gas.

\subsection{Exchange field perpendicular to the plane of 2D gas}

The case of exchange field perpendicular to the plane of two-dimensional gas, i.e. $H_{z} \neq 0$ and $H_{x} = H_{y} = 0$, corresponds either to ferromagnetic LAO/STO layers or to antiferromagnetic system with uncompensated interface.

\begin{figure*}[t]
 	\includegraphics[width=2.0\columnwidth]{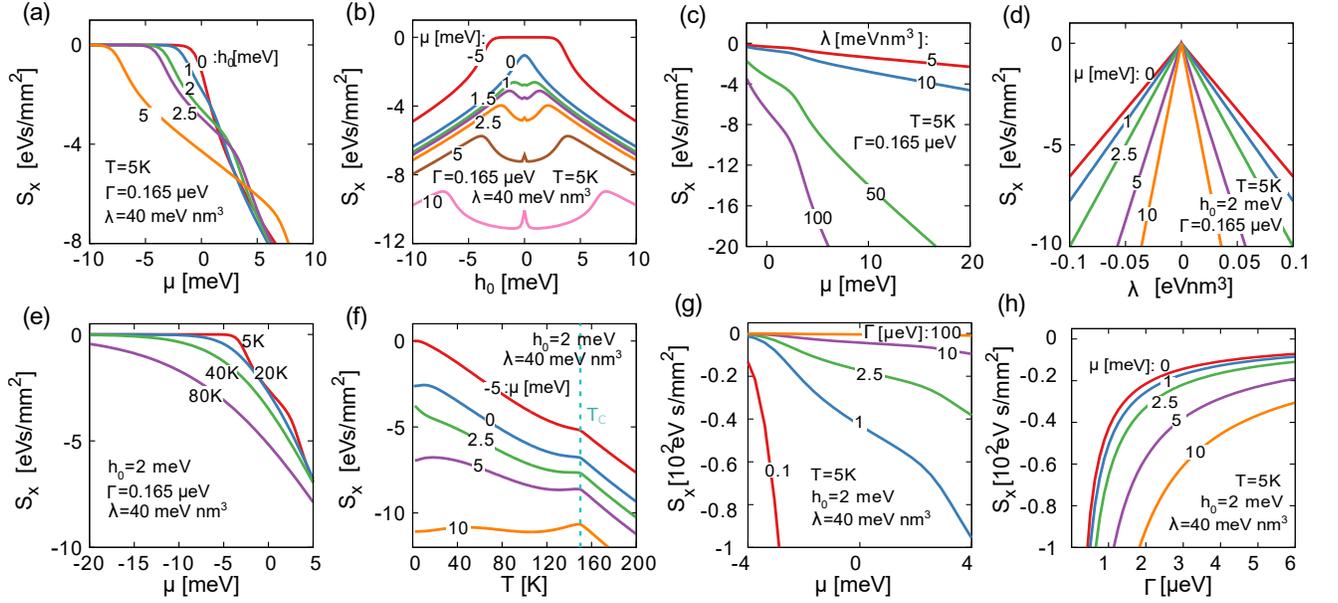}
 	\caption{The x-component of current-induced spin polarization for exchange field normal to the plane of 2DEG (2DHG) shown as a function of chemical potential (a,c,e,g), exchange field $h_0$ (b), spin-orbit coupling parameter $\lambda$ (d), temperature T (f), and relaxation rate $\Gamma$ (h) for fixed parameters as indicated.
 		 		Other parameters are the same as in Fig.2.}
 	\label{fig:fig4}
 \end{figure*}

\begin{figure}[t]
	\includegraphics[width=1.0\columnwidth]{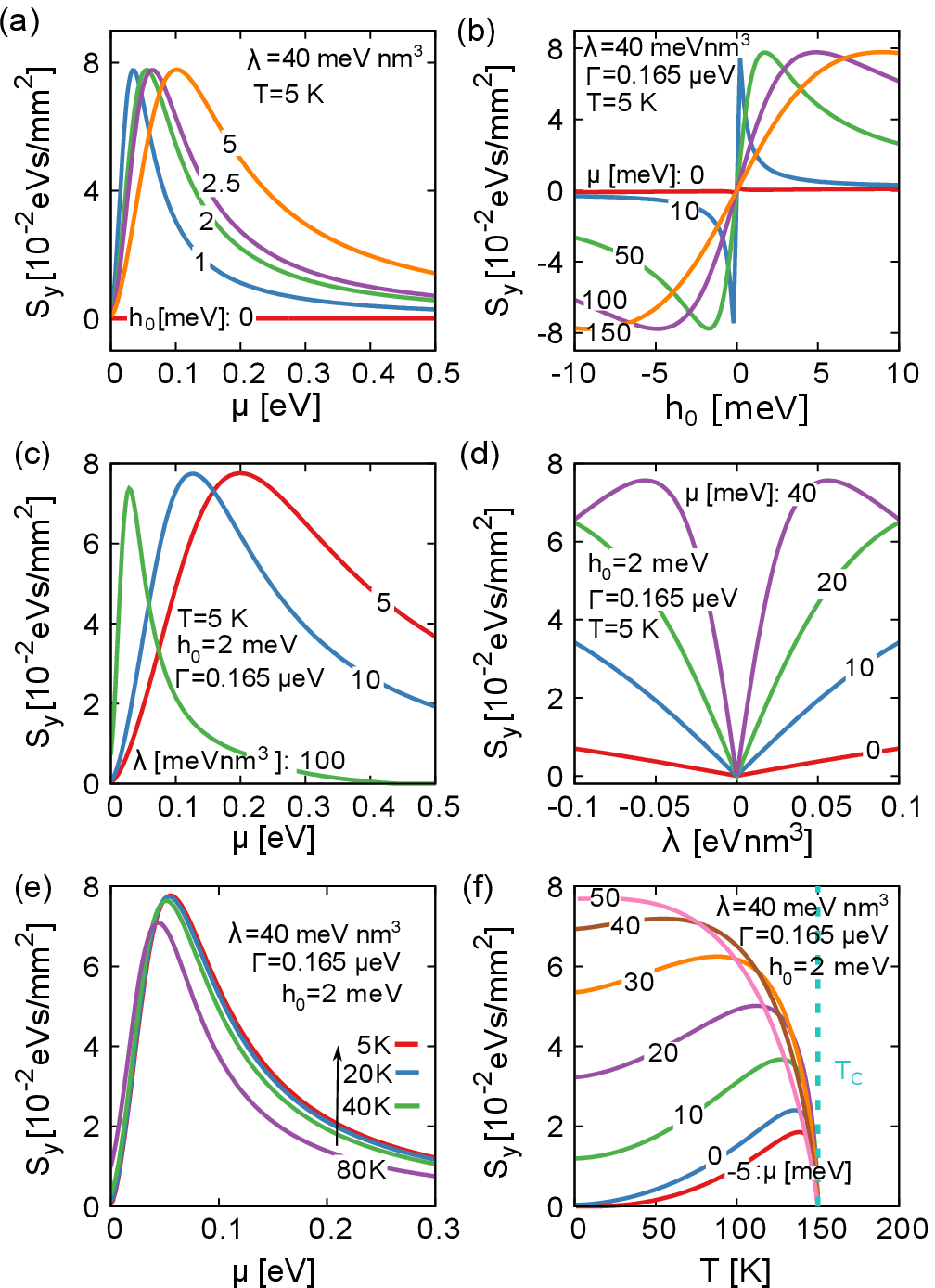}
	\caption{The y-component of current-induced spin polarization for exchange field normal to the plane of 2DEG (2DHG)  shown as a function of chemical potential (a,c,e), exchange field $h_0$ (b), spin-orbit coupling parameter $\lambda$ (d), and temperature T (f)  for fixed parameters as indicated. Other parameters
				as in Fig.2.}
	\label{fig:fig5}
\end{figure}

 The non-equilibrium spin polarization has then two components, namely the $S_x$ component (which remains also finite for zero exchange field), and the $S_y$ component that is absent in the limit of zero exchange field. The numerical results are presented in Figs.~\ref{fig:fig4} and ~\ref{fig:fig5}.

The $S_x$ component of spin polarization is only weakly modified by the perpendicular exchange field which introduces, e.g., small nonlinearities in the dependence of $S_{x}$ on the chemical potential,  clearly seen in Figs.~\ref{fig:fig4}(a),(c),(e) and (g). These nonlinearities can be attributed to the presence of an energy gap between the subbands.
For small values of chemical potential, the component  $S_{x}$ varies monotonically  with increasing magnitude of exchange field, whereas for larger values of $\mu$ this behaviour is nonmonotonous, see Fig.~\ref{fig:fig4}(b).
In the latter case
we observe a local minimum which appears when the Fermi level crosses the bottom edge of the higher subband. For larger values of $|h_{0}|$
only one subband is occupied and the spin polarization increases with a further increase in the exchange field, see Fig.~\ref{fig:fig4}(b).
The $x$-component of spin polarization changes linearly with $\lambda$, as shown explicitly in Fig.~\ref{fig:fig4}(d).
 The temperature dependence, Fig.~\ref{fig:fig4}(f), is similar to that in the absence of exchange field. However, some kinks are well pronounced when the temperature approaches the Curie temperature and the system becomes nonmagnetic.

As the $x$-component of the current-induced spin polarization is only slightly modified by the perpendicular exchange field, the $y$-component of spin polarization appears solely when the exchange field is nonzero. For the perpendicular exchange field, Eq.(\ref{sy}) leads to the following expression for $S_y$:
\begin{eqnarray}
\label{syHz}
S_{y} = e E_{y} \hbar s_{1} \int \frac{d^{2} \mathbf{k}}{(2\pi)^{2}} \frac{9 H_{z} k^{4} \lambda}{4 n^{3}} \left[ f(E_{+}) - f(E_{-})\right]\nonumber\\
- e E_{y} \hbar s_{1} \int \frac{d^{2} \mathbf{k}}{(2\pi)^{2}} \frac{9 H_{z} k^{4} \lambda}{4 n^{2}} \frac{\Gamma^{2}}{n^{2} + \Gamma^{2}} \left[ f'(E_{+}) + f'(E_{-})\right]\nonumber\\
\end{eqnarray}	
where $n$ is reduced to the following  form: $n=\sqrt{(3H_z)^2+(2\lambda k^3)^2}/2$.

Two important features of the spin polarization follow from the above expression. First, the $S_{y}$ is linear with respect to the parameter $s_{1}$ (which determines the off-diagonal elements of spin operators). This means that even though the contributions associated  with $s_{1}$ lead only to a small correction to the $x$-component of spin polarization (see the discussion in the preceding section), they are responsible for the $y$-component of the spin polarization. Therefore, one cannot ignore the terms related to $s_{1}$ if one wants to describe  properly the physics of spin polarization in the model under consideration. Second, the $S_{y}$ component is related to the topological properties of the system and may be expressed in terms of the Berry curvature. The above expression is therefore, in its main part, robust against scattering on impurities. For long relaxation time ($\Gamma \to 0$), the formula (\ref{syHz}) reduces to:
	\begin{eqnarray}
	\label{SyforHz}
	S_{y} = e E_{y} \hbar s_{1} \int \frac{d^{2} \mathbf{k}}{(2\pi)^{2}} \frac{9 H_{z} k^{4} \lambda}{4 n^{3}} \left[ f(E_{+}) - f(E_{-})\right].
	\end{eqnarray}	
The Berry curvature $\bB_{j}$ for the $j$-th subband is defined by the Berry connection, $\bA_{j}(\mathbf{k}) = \mathrm{i} \langle \Psi_{j}| \nabla_{\mathbf{k}} | \Psi_{j} \rangle$, as
\begin{equation}
\bB_{j} = \nabla_{\mathbf{k}} \times \bA_{j}(\mathbf{k}).
\end{equation}
For two-dimensional systems confined in the $xy$-plane only $z$ component of Berry curvature is present. For k-cubed Rashba gas with exchange field oriented perpendicularly to plane one obtains:
\begin{eqnarray}
\label{BerryCurvature}
\mathcal{B}^{z}_{\pm} = \pm \frac{54 H_{z} \lambda^{2} k^{4}}{(2n)^{3}}.
\end{eqnarray}
Thus, combining (\ref{BerryCurvature}) with (\ref{SyforHz}) we get the following  simple expression for the y-component of spin polarization:
\begin{eqnarray}
S_{y} = e \hbar E_{y} \frac{s_{1}}{3 \lambda} \sum_{j = \pm} \int \frac{d^{2} \mathbf{k}}{(2\pi)^{2}} \mathcal{B}_{j}^{z} f(E_{j}).
\end{eqnarray}

Variation of the $y$-component of spin polarization with $\mu$, $h_{0}$, $\lambda$ and T is presented in Fig.~\ref{fig:fig5}.
When the Fermi level increases starting from small values, the $S_y$ component also increases until it reaches its  maximal value that depends mainly on
the $s_1$ parameter. Then, the $S_y$ component decreases with a further increase in the Fermi energy, as shown in Figs.~\ref{fig:fig5}(a) and (b). Furthermore, the maximum in $S_y$ shifts to higher Fermi levels with increasing exchange energy. It is worth to note that the $S_y$ component can change its sign when the magnetization is reversed.
For relatively small Rashba spin-orbit coupling strength and small values of $\mu$, the $S_y$ component increases monotonously with $\lambda$, as shown in Figs.~\ref{fig:fig5}(c) and (d). However, for larger values of $\mu$, the $S_y$ component initially increases with $\lambda$ and then upon reaching a maximum it decreases with a further increase in $\lambda$. This behaviour is different from that found for the $S_x$ component.
Since the $S_y$ component is strongly dependent on the exchange field, it is also highly sensitive to changes in temperature, as shown in Figs.~\ref{fig:fig5}(e) and (f). For small Fermi levels the maximal value of $S_y$ occurs when $T$ approaches Curie temperature $T_\text{C}=150~\textrm{K}$ and then it quickly disappears when the exchange field vanishes, i.e. when $H_z(T=T_\text{C})=0$. For higher Fermi levels, the maximal value occurs at low temperatures.

\subsection{Exchange field in the plane of 2D gas}

\begin{figure*}[t]
	\includegraphics[width=\textwidth]{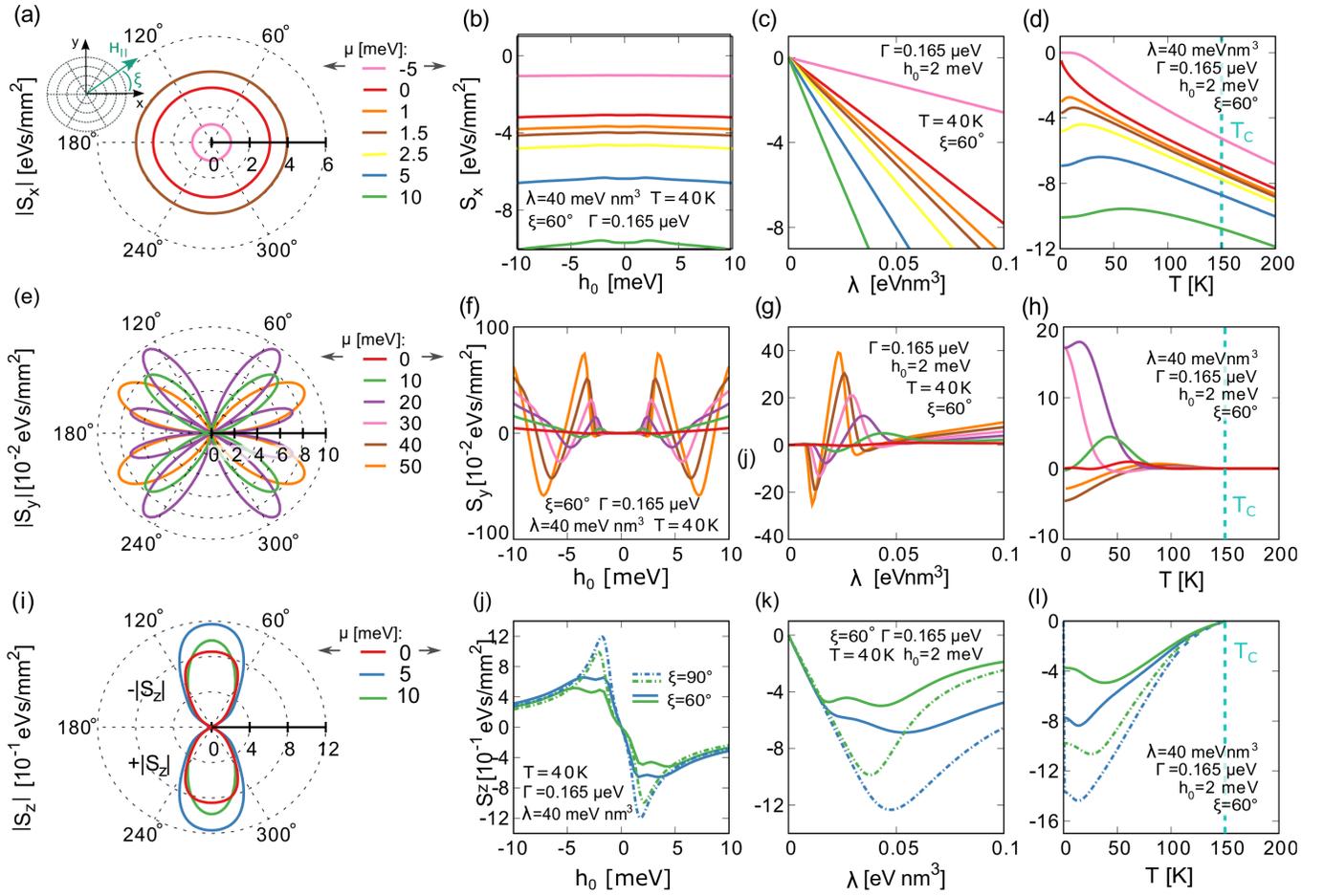}
	\caption{Current-induced spin polarization for exchange field in plane of 2DEG (2DHG). The components $S_x$, (a), $S_y$, (e), and $S_z$, (i), as a function of the angle $\xi$ describing orientation of the in-plane exchange field and for  indicated values of electrochemical potential $\mu$. The component $S_x$, $S_y$, and $S_z$ as a function of exchange field $h_0$, (b), (f), (j); Rashba coupling strength $\lambda$, (c), (g), (k); and temperature $T$, (d), (h), (l), respectively.
			Other parameters as in Fig.2.}
	\label{fig:fig6}
\end{figure*}

Consider now the case when the exchange field is in the plane of the 2DEG, i.e. when $H_z=0$ and $H_x, H_y\neq 0$.
Similarly as in the case described above, the $x$-component of spin polarization is only weakly modified by the  in-plane exchange field -- see Fig. ~\ref{fig:fig6}(a-d).
On the other hand, when the exchange field is in the plane of two-dimensional gas, both $y$ and $z$ components of the spin polarization can occur and significantly depend on the orientation and strength of the exchange field.  For example, the  $S_y$ component (see Fig.~\ref{fig:fig6}(e)) is absent when the in-plane field is parallel to either $x$ or $y$ axis (i.e. when $\xi=0^\circ$ or $\xi=90^\circ$), whereas the $S_{z}$ component vanishes for the in-plane field oriented along the $x$ axis and takes its maximal value for exchange field oriented along the $y$ axis, see Fig.~\ref{fig:fig6}(i). The $x$ and $y$ components are both nonzero  when the in-plane exchange field is aligned  between the $x$ and $y$ axes. The specific positions of the maxima in $S_{y}$  depend on the Fermi level.
Both $S_{y}$ and $S_{z}$ components, however,  are one to two orders of magnitude smaller than the $S_{x}$ component.

Behavior of spin polarization presented in Fig.~\ref{fig:fig6}  indicates on a strong interplay between the effective field induced by spin-orbit coupling and the in-plane exchange field. Dependence of the $S_{y}$ component on the exchange field and Rashba coupling parameter is presented in Figs.~\ref{fig:fig6} (f) and (g), for $\xi = 60^\circ$. This component behaves symmetrically with respect to the magnetization reversal, and its sign can be changed by  tuning  magnitude of the exchange field or spin-orbit coupling strength. The temperature-dependence of $S_y$, shown in Fig.~\ref{fig:fig6}(h), indicates that relatively low temperatures are necessary to have  remarkable  values of $S_y$ and that the $S_y$ component vanishes when $T$ approaches the Curie temperature.

The $S_{z}$ component, in turn,  is antisymmetric with respect of the sign reversal of the in-plane exchange field.
A nonzero value of  $S_{z}$ means that the vector of spin polarization is tilted out of the plane of 2D gas.
In Figs. ~\ref{fig:fig6}(j)-(k) the dependence of $z$-component of spin polarization on the magnitude of  exchange field and Rashba coupling parameter is presented for two orientations of the field, i.e. for  $\xi=90^\circ$ and $\xi=60^\circ$.  The maximal values of $S_z$ occur for the exchange field oriented parallel to the electric field, that is for $\xi=90^\circ$. Moreover the $S_z$ component is larger for higher values of chemical potential. For $\xi=90^\circ$, the $|S_z|$ component displays only one peak in the dependence on $h_0$, see Fig.~\ref{fig:fig6}(j). For $\xi=60^\circ$ and, more generally, $\xi\in (0,90^\circ)$, the $|S_z|$ curve displays two peaks. This might be attributed to the anisotropy introduced by the in-plane field and greater separation in the $\mathbf{k}$-vector space of the $E_{-}$ and $E_{+}$ states for $\xi=60^\circ$ than for $\xi=90^\circ$. Similar behaviour is visible in Fig.~\ref{fig:fig6}(k), where the $S_z$ component is shown as a function of Rashba spin-orbit coupling strength $\lambda$. Similarly as in  the case of $S_y$ component, the $S_z$ component is remarkable for low temperatures and disappears when the temperature approaches the Curie temperature, where the exchange field vanishes.

\section{Summary}

We  presented  a detailed study of  the current-induced spin polarization in two-dimensional electron gas with
an isotropic $k$-cubed Rashba spin-orbital coupling. The model under consideration is useful for understanding the nonequilibrium spin polarization and spin dynamics in some p-doped semiconductor quantum wells, as well as  in electron gases at the interfaces of oxides perovskites.  We have shown that the contribution related to the parameter $s_1$ should not be omitted.
This contribution in a nonmagnetic case modifies the spin polarization by about 10\%, however in the magnetic case it is responsible for the components that are absent in the limit of zero exchange field.
We  also discussed briefly  the relation of some terms in the spin polarization with the Berry curvature.
Generally, one can expect that 
the induced spin polarization in a magnetic system leads to a torque  on the local magnetization, which in turn can  modify the spin dynamics.

\begin{acknowledgments}
	This work was partially supported by  the Polish Ministry of Science and Higher Education through a research project ’Iuventus Plus’ in years 2015-2017 (project No. 0083/IP3/2015/73) and by the German Research Foundation (DFG) under SFB 762. A.D. acknowledges the support from the Foundation for Polish Science (FNP).
\end{acknowledgments}

%%%%%%%%%%%%%%%%%%%%%%%%%%%

\end{document}